# Ivermectin and Doxycycline Combination as a Promising Drug Candidate Against SARS-CoV-2 Infection: A Computational Study

Meenakshi Rana[1], Pooja[2], Papia Chowdhury[2]*

[1]Department of Physics, School of Sciences, Uttarakhand Open University, Haldwani, 263139, Uttarakhand, India

[2]Department of Physics and Materials Science and Engineering, Jaypee Institute of Information Technology, Noida 201309, Uttar Pradesh, India.

*Corresponding author: papia.chowdhury@jiit.ac.in

## Abstract

In the present study, we have described how by using molecular docking and molecular dynamic (MD) simulation studies the combination drug of ivermectin and doxycycline can be used as a potential inhibitor for SARS-CoV-2 virus. In lieu of unavailability of specific cure of COVID-19 till now various possibilities for individual and combination drugs have been explored by the medical practitioners/scientists for the remedial purpose of CoV-2 infections. 3CL$^{pro}$ is the main protease of SARS-CoV-2 virus which plays an essential role in mediating viral replication in the human body. 3CL$^{pro}$ protein can serve as an attractive drug target. In this work, we have studied drug: 3CL$^{pro}$ interactions by in silico molecular docking and MD simulation approaches. Common and easily available antiviral drugs ivermectin, doxycycline and their combination have been proved their valid candidature to be used as potential drug candidates against SARS-CoV-2 infections.

# 1. Introduction:

In the year 2020, the COVID-19 disease has spread globally and it has become an ongoing pandemic. Reported by the World Health Organization (WHO), due to this pandemic disease, more than 35,659,007 numbers of active patients with 1,044,269 people have already died till 10 October 2020 (https://covid19.who.int/). WHO declared the COVID-19 as a global health emergency. This disease is caused by a member of the coronavirus family [1]. Coronavirus was first found in 1930 in domestic poultry [2]. After that they were identified as causing several diseases in humans such as; respiratory illness, neurological, liver diseases, etc. [3]. Till now seven categories of this virus were identified. Among the seven categories of coronavirus, four causes only common cold with mild symptoms and in very rare cases pneumonia, respiratory infections in infants and older people [4]. The other three categories are Severe Acute Respiratory Syndrome Coronavirus (SARS-CoV) [5], Middle East Respiratory Syndrome Coronavirus (MERS-CoV) [6] and lastly the new one known as SARS-CoV-2 [7] identified in 2003, 2012 and 2019 respectively. The international committee on taxonomy of viruses declared this new novel coronavirus as SARS-CoV-2 [8]. The SARS-CoV-2 is a single-stranded RNA virus and belongs to the Coronaviridae family having genome sequences of 79.5% sequence matching [9,10]. This shows that bats may be the carrier of this virus. The uniqueness of this virus is the presence of spike glycoproteins on its surface which gives a crown-like appearance of the virus structure. The crown-like spike protein surface of this virus can be easily visible with the help of electron microscopes. These spike proteins are a very significant part of SARS-CoV-2 [11] virus as they can easily interact with the human proteins which coats the inside of the nose and the cells of lungs. The interaction of spike protein and human protein causes change in spike protein of CoV-2 shape and causes the human receptor cell to swallow up the virus. Through the receptor binding domain (RBD), glycoproteins of the viruses start binding and entering to the host cells. The key receptor for SARS-CoV-2 in humans is angiotensin converting enzyme 2 (ACE2) [12]. After entering the host cell, different human protease like airway trypsin-like protease (HAT), cathepsins and trans membrane protease serine 2 (TMPRSS2) divide the glycoproteins of the virus and so the conformational alteration of the virus structure occurs. From this phase the transformed virus replicates itself very fastly through some cyclic processes [12] and starts infecting the neighboring cells like lung, heart, brain cells and many others. From studies, scientists showed that the spike glycoproteins of coronavirus attach on the cell surface of the ACE2 receptor in the human body and allows the virus's genetic material to enter the human cell [13]. Virus's genetic material proceeds to hijack the metabolism of the cell and help the virus to divide.

The main symptoms of this disease are fever, tiredness, dry cough. Other symptoms include shortness of breath, body pains, soreness in the throat and a small number of people reported diarrhea, running nose [14]. At the beginning these symptoms appear generally in mild form and gradually increase afterwards. The first infected patient by this virus was detected in Wuhan, China in December, 2019 [15,16]. The highly contagious nature of this virus causes fast spreading of the diseases and became an ongoing pandemic virus spread globally. The spreading of virus occurs by the close contact along with the droplets spilled during talking, coughing and sneezing from the infected person [17]. Research works show that the chance of being infected by COVID-19 reduces by maintaining "social/physical" distancing along with proper hand-hygiene. Though there are many predictions about the airborne transmission of this disease, no scientifically valid evidence is available till now [18]. Depending upon the age and immunity of the person the symptoms likely to appear within two to fourteen days after infection with the virus [19]. Mortality increases with people aged over 60 years and having diseases such as hypertension, immune-weakening medications, diabetes, cardiovascular disease, chronic respiratory disease and cancer etc. Elderly people accounted for 42% of total fatalities and people with several diseases accounted for 78% of total deaths [20]. However, very rare and mild with about 2.4% of the total reported cases have been reported in children (below age 19 years) [20].

To overcome this disease the whole world is in a race to find vaccines/drugs to attack this virus. Through clinical trials around 200 drugs and vaccines (approved by Food and Drug Administration). Covaxin, INO-4800, mRNA-1273, NVX-CoV2373, BBV152 etc. are some candidate vaccines that are currently under trials for COVID-19 [21]. Similarly examples of some FDA approved drugs for COVID-19 are atazanavir, remdesivir, ritonavir, lopinavir, chloroquine, hydroxychloroquine (HCQ), cyclosporin, favipiravir etc. [22 -24]. Now according to most common treatment protocols since there is no detected and approved drug for COVID-19, patients with severe COVID-19 symptoms are usually treated by different purposed antiviral drugs as trial basis. Most of the above-mentioned drugs are usually antiviral in nature and are used for various viral diseases like: HIV medication, influenza, MERS and SARS diseases or for enhancing the immune system of human life [25-27]. Nowadays to identify potential drugs for various diseases, the concept of drug repurposing is widely used. Drug repurposing is an approach to find out the new uses for already available drugs that are originally developed for specific diseases [28]. Drug repurposing process has already proved to be very effective since many drugs have multiple protein targets and genetic factors; molecular pathways which can be shared by diverse diseases. For many years repurposing of drugs have been used such as favipiravir drug used for influenza virus, sofosbuvir drug used for hepatitis C virus have a strong repurposing prospective against Zika and Ebola [29], drugs oseltamivir, lopinavir, nelfinavir, atazanavir and ritonavir have been used for the treatment SARS and MERS [30,31]. But these drugs have their own toxicity related issues. On the other hand, some immunomodulatory plasma-based therapies are in use. Some food nutrients, herbal medicines having antiviral and immunity building properties are considered as an alternative of COVID-19 therapies [32,33]. In the same way, a repurposing of combination drugs with ribavirin, lopinavir, and ritonavir have already been anticipated for the COVID-19 patients [34]. Lopinavir and ritonavir combination are already in use for HIV treatment. However, till now no specific drugs and vaccines to combat against the COVID-19 have been discovered. So there is an urgent and strong requirement for a newly invented drug/repurposed drug/combination drug to fight the disease.

A combination drug includes two or more than two active ingredients mixed in a single dose form. For many years combination of drugs has been used for treating diseases such as aspirin/paracetamol/caffeine combination (Excedrin) is used for the treatment of headache and migraine [35], Carbidopa/levodopa/entacapone is used for the treatment of Parkinson's disease [36], and indacaterol/mometasone, used for the treatment of asthma [37]. Combination drug therapy is applied for many diseases such as: tuberculosis, leprosy, cancer, bacterial infections, malaria, and for many viral diseases like influenza, HIV/AIDS etc. [38]. Recently two combination drugs of Nitazoxanide/azithromycin [39] and another combination drug: lopinavir/oseltamivir/ritonavir are [40] being largely in use by medical practitioners to fight against SARS-CoV-2 infections. There are several advantages to the combination of drugs. They are increased action of drugs and efficiency, increase the efficiency of the therapeutic effect, reduced cost and side effects. However, combinations of drugs also include some disadvantages. Dose must be given in some fixed ratio otherwise mismatched pharmacokinetics may increase severe toxicity effects. Most important part of combination drugs is that repurposing of common available drugs may reduce cost, time of action and risk factor. Though several clinical trials are underway to identify drugs against SARS-CoV-2, but still currently there is availability of single approved drugs or vaccines. Urgent requirement of cure of current medical emergencies due to COVID-19 motivated us to investigate the possibility of inhibition of SARS-CoV-2 by using some repurposing of combination drugs: ivermectin and doxycycline.

In the present paper, we have described how the combination drug of ivermectin and doxycycline, can be used as a potential SARS-CoV-2 M$^{pro}$ inhibitor. For several years ivermectin ($C_{48}H_{72}O_{14}$) is used to treat many infectious diseases in mammals. It is orally prescribed and has a low toxicity profile. Ivermectin has a broad-spectrum drug and possesses numerous effects on parasites, nematodes, arthropods, mycobacteria, flavivirus, and mammals [41]. By specifically targeting its NS3 helicase, it was also used to cure Japanese encephalitis virus (JEV) and yellow fever virus (YFV) [42,43]. In the late 1970s, ivermectin was first known and in 1981 permitted for the use of animals [44]. Doxycycline is an antibiotic drug and used to treat the infections caused by bacteria. Doxycycline ($C_{22}H_{24}N_2O_8$) synthetically derived from oxytetracycline. This drug is a second-generation tetracycline, which is readily absorbed and bound to plasma proteins. It is mainly used for the treatment of pneumonia, respiratory tract infections, rocky mountain spotted fever, typhus fever and the typhus group, rickettsialpox, tick fevers, and urinary tract infections etc. It is also used to prevent malaria. Normally in the market it comes as a capsule, tablet, and suspension (liquid) to take orally. We have performed molecular docking and molecular dynamics (MD) simulations to understand the interaction mechanism of the proposed drugs for COVID-19.We hope that this work will provide other researchers with an important investigation way to identify new COVID-19 treatment.

## 2. Materials and Methods:

### 2.1. Protein structure preparation

Coronavirus possesses a number of polyproteins (structural and nonstructural). Among them 3CL$^{pro}$ is a key CoV enzyme which plays an important role in mediating viral replication and transcription with the help of its glycoprotein. To rapidly discover the targeted drugs for clinical use, researchers focused on identifying drug leads that target 3CL$^{pro}$ protein of SARS-CoV-2 as it plays an important role for viral replication and transcription. In the present work, we have used one of 3CL$^{pro}$ proteases of CoV-2 virus in a complex with an inhibitor N3 (PDBID: 6LU7) [45,46] as the target protein. 6LU7 can be proved to be an attractive target for designing effective drugs for COVID-19. We have chosen 6LU7 protein for checking the inhibiting and binding properties of it with the ivermectin and doxycycline drugs. The structure of SARS-CoV-2 protease (6LU7.pdb) was used as a receptor and retrieved from Protein Data Bank (http://www.rcsb.org/) [45,47] and are shown in Figure 1 (a). We have removed water and hydrogen from it. All the existing properties of the proteins are described in Table 1. For the preparation of protein, we have used Auto Dock and MG Tools of AutoDock Vina software [48]. At first existing lead components, water molecules and ions have been removed from it. Later the process of cleaning has been done. We have calculated the Gasteiger charges of protein structures and after that polar hydrogen have been introduced. Then the non-polar bonds were merged and rotatable bonds were defined. Finally, by using Discovery studio 2020 [49] the intrinsic ligands were detached from the protein molecules and the final protein molecule was saved in the PDB format (Figure 2 a).

**Figure 1. a)** Structure of receptor protein (6LU7). **b)** Structure of ivermectin **c)** Structure of doxycycline (from Protein data bank and Gauss view). In the figure red color: oxygen atom, blue: nitrogen atom, gray colour: carbon atom.

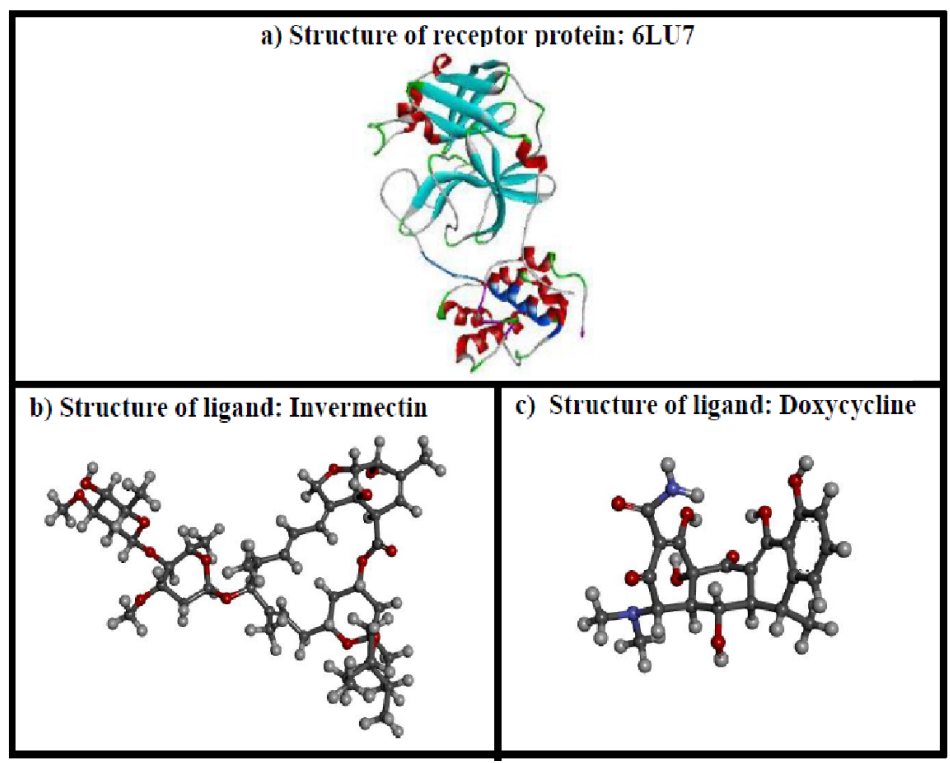

For target protein by visualizing the dihedral angles ψ against φ of amino acid residues, Ramachandran plots have been drawn (Figure 2b). It predicted permissible and disfavored values of ψ and φ. Ramachandran plots for 6LU7 have been shown in the figure 2b, plot specifies localization on the chain residues, which represent the quality of the protein structure means efficient and accurate docking potential.

**Figure 2. a)** Target variable viral proteins (6LU7) SARS-CoV-2 protease enzyme as receptor and **b)** Ramachandran plot for the receptor protein.

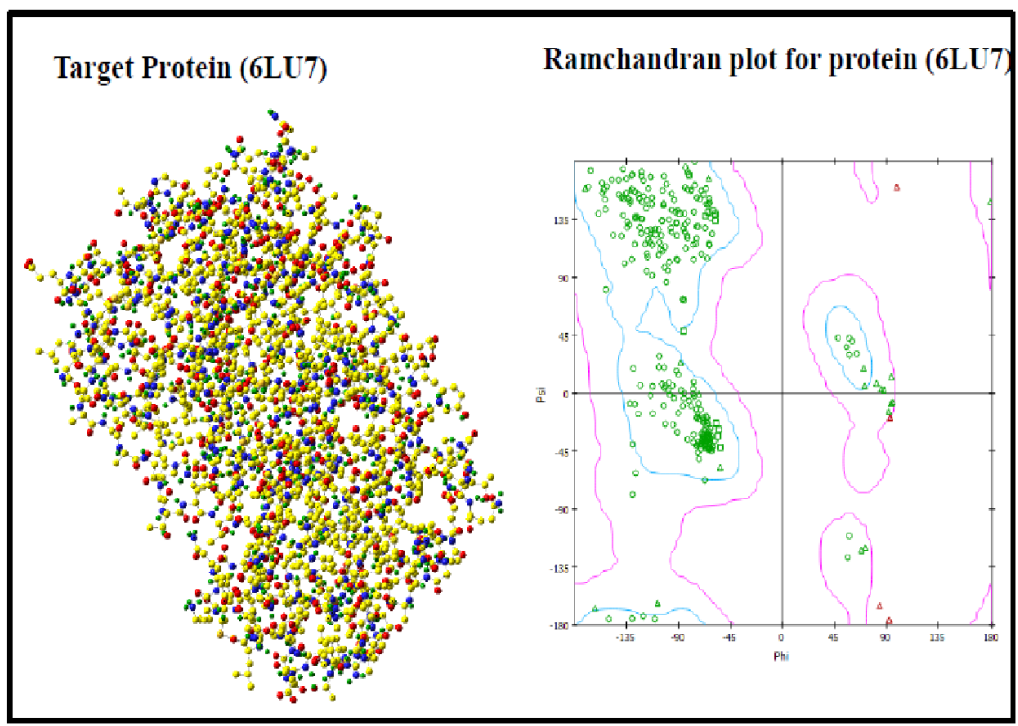

## *2.3. Ligand drug molecules preparations*

Structures of drug molecules were drawn by using the Gaussian 09 program [50]. All the drugs optimization of geometries were carried out with Hartree fock and STO 3G basis set. Gauss View 5 molecular visualization program was used for visualizing the optimized structure [51]. Open Babel software has been used to prepare 3D coordinates. ADME-T properties of molecules were identified using Organic chemistry portal (http://www.organic-chemistry.org/prog), a web based application for predicting *in-silico* ADME-T property. Protein–ligand interactive visualization and analysis was carried out in AutoDock 4.2 software on Windows 7 (64-bit).

For the present work, we have selected two potential ligand drugs: ivermectin ($C_{48}H_{72}O_{14}$) and doxycycline ($C_{22}H_{24}N_2O_8$). Detail structures of these molecules were downloaded from Drug Bank in pdb format (Figure 1 and Table 1). Different chemical, physical, drug likeness and pharmacokinetics properties obtained from SWISS ADME and shown in Table 1. Both the proposed drug molecules have molecular weight less than 875 gm/mol and topological polar surface area (TPSA) values less than 180 $Å^2$ (Table 1). All drug molecules have H-bond donors ≥6, H-bond acceptor ≥14 and have low synthetic accessibility count, this suggests that they can be synthesized easily. Though these drugs violate some drug likeness properties, still the availability of these drugs in the drug industry motivates us to consider these as potential inhibitors. Molecular docking study requires the ligand file in pdbqt format. AutoDock Tools 1.5.6 [52] have been used to save ligands in pdbqt format.

*2.3. Methods: Molecular docking and Molecular dynamics simulations*

To predict the target and drug interactions, molecular docking is commonly used in simulation. It minimizes the energy and calculates the binding energy of the interactions. In the molecular docking simulation, we normally make out the best pose of the ligand towards the receptor protein with the help of scoring functions [53]. Molecular docking can show the possibility of any biochemical reaction or whether a drug is docked with the receptor protein or not. The AutoDock vina with the best fitted parameters binding modes – 9, exhaustiveness – 8, applied maximum energy difference – 3 kcal/mol and Grid box center with x, y, and z coordinate of residue position of the protein is used for docking purpose [48]. AutoDock Tools [48] have been used for saving the target protein in pdbqt format. The criteria for choosing the best position from the docked 9 modes is the maximum nonbonded interaction, lower binding affinity (kcal/mol), dipole moment (Debye), dreiding energy and inhibition constant. Best ligand: protein pose is identified by knowing the types (H-bonds, hydrophobic bonds) and number of bonding between them. The drug which makes the maximum number of bonds with the target protein mostly shows better complex formation. For analyzing and visualizing non-bonded hydrogen bonds for different output poses, Discovery Studio visualizer 2020 version 20.1.0.19295 [49] have been used. After the analysis of individual docking, sequential docking is performed. For sequential docking, the grid box coordinates were set to the particular binding region of each drug with default grid spacing. In the procedure of sequential docking, the first ligand is docked and the complex is saved out as a single file, where the first ligand is considered part of the receptor. Docking is then carried out on this complex with the second ligand. The structural dynamics of receptor and inhibitor interaction and thermodynamics stability of ligand: protein have been investigated with the help of Linux based platform "GROMACS 5.1 Package" [54], Different thermodynamic parameters like temperature (T), density (D), potential energy ($E_{pot}$), root mean square deviation (RMSD) for backbone, root mean square fluctuation (RMSF) for protein $C_\alpha$, solvent accessible surface area (SASA), intermolecular hydrogen bonds, interaction energies (ΔG) of the protein and drug complex have been find out with CHARMM36 all atom [55] and GROMOS43A2 force fields [56]. In aqueous solution simulations have been performed using the water model: TIP3P. For solvation process protein in bare state, protein: ligand complex were solvated in a cubic box, with a buffer distance of 10 Å and volume as 893,000 Å$^3$. For electrically neutralizing the system four Na$^+$ ions have been added. Then we minimize the energy in the vacuum. For energy minimization 50000 iterations have been taken. In the present work, we have observed that within 10 ps of the time scale all the complex formation reached stability so for the study we have restricted our simulation upto 10 ns. Number of particles (N), volume (V), and temperature (T) were constant under the 1 atmosphere pressure and 298K temperature. We have used Lennard-Jones and Coulomb short range interaction for the nonbonded interactions. Graphical tool Origin pro has been used to study the simulated results.

"Molecular Mechanics Poisson-Boltzmann Surface Area" (MMPBSA) method [57] have been used for calculating the interaction free energies (ΔG$_{bind}$) of the protein: drug complex. ΔG$_{bind}$ calculation usually begins after the MD simulation of the complex using the single trajectory approach. ΔG$_{bind}$ in the aqueous solvent, for the bound protein: ligand complex can be given as:

$$\Delta G_{bind,aqu} = \Delta H - T\Delta S \approx \Delta E_{MM} + \Delta G_{bind,solv} - T\Delta S \ldots\ldots\ldots\ldots\ldots..(1)$$

$$\Delta E_{MM} = \Delta E_{covalent} + \Delta E_{electrostatic} + \Delta E_{vander\ waals} \ldots.(2)$$

$$\Delta E_{covalent} = \Delta E_{bond} + \Delta E_{angle} + \Delta E_{torsion} \ldots\ldots\ldots\ldots\ldots(3)$$

$$\Delta G_{bind,solv} = \Delta G_{polar} + \Delta G_{nonpolar} \ldots\ldots\ldots\ldots\ldots\ldots\ldots\ldots(4)$$

Where, $\Delta E_{MM}$ is the molecular mechanical energy changes in gas phase and is the sum of covalent $\Delta E_{covalent}$, electrostatic ($\Delta E_{electrostatic}$), and van der Waals energy ($\Delta E_{vander\ waals}$) changes. Covalent energy is the combination of bond angle and torsion and $\Delta G_{bind,solv}$ is separated into its polar and nonpolar contributions. $\Delta G_{bind,solv}$ is solvation free energy change and -TΔS conformational energy change due to binding.

*2.4. Computational facility*

MD simulations and corresponding energy calculations have been computed using HP Intel Core i5 - 1035G1 CPU and 8 GB of RAM with Intel UHD Graphics and a 512 GB SSD.

## 3. Results and discussion

### 3.1. Individual docking of drugs against SARS-CoV-2 protease

Ivermectin's potential application for the treatment of various diseases in humans was confirmed a few years later. William C. Campbell and Satoshi Ōmura received the 2015 Nobel prize in physiology or medicine for the discovery and development of this drug [44,58,59]. Ivermectin has a broad-spectrum drug and possesses numerous effects on parasites, nematodes, arthropods, mycobacteria, flavivirus, and mammals [60]. By specifically targeting its NS3 helicase, it was also used to cure Japanese encephalitis virus (JEV) and yellow fever virus (YFV) [43,61]. Along this, it is also able to increase the immune system. Doxycycline is an antibiotic drug and used to treat the infections caused by bacteria. Doxycycline ($C_{22}H_{24}N_2O_8$) synthetically derived from oxytetracycline. This drug is a second-generation tetracycline, which is readily absorbed and bound to plasma proteins. It is mainly used for the treatment of pneumonia, respiratory tract infections, rocky mountain spotted fever, typhus fever, rickettsialpox, tick fevers, and urinary tract infections etc. It is also used to prevent malaria. Normally in the market it comes as a capsule, tablet, and suspension (liquid) to take orally.

In the present work, ivermectin and doxycycline drugs were docked to SARS-CoV- 2 main protease (3CL$^{pro}$) protein (6LU7). Ivermectin and doxycycline drugs confirm the Ro5 and other drug likeness rules etc. Hence, we have shown their strong application as potential drugs reaching the market (Table 1).

**Table 1.** Molecular configuration and drug likeness properties of proposed ligand drug molecules for COVID-19 by SWISS ADME data.

| Pub Chem CID | **6321424** | **54671203** |
|---|---|---|
| Name of Ligand | Ivermectin | Doxycycline |
| **Physiochemical Properties** | | |
| **Molecular Formula** | $C_{48}H_{74}O_{14}$ | $C_{22}H_{24}N_2O_8$ |
| Molecular Weight (g/mol) | 875.09 g/mol | 444.43 g/mol |
| Hydrogen Bond Donor Count | 3 | 6 |
| Hydrogen Bond Acceptor Count | 14 | 9 |
| Rotatable Bond Count | 8 | 2 |
| Topological Polar Surface Area | 170.06 Å² | 181.62 Å² |
| Heavy Atom Count | 62 | 32 |
| Formal Charge | 0.81 | 0.41 |
| Molar Refractivity | 230.77 | 110.91 |

| | Lipophilicity | |
|---|---|---|
| Log $P_{o/w}$ (iLOGP) | 5.86 | 1.93 |
| Log $P_{o/w}$ (XLOGP3) | 6.34 | 0.54 |
| Log $P_{o/w}$ (WLOGP) | 5.60 | -0.50 |
| Log $P_{o/w}$ (MLOGP) | 1.25 | -2.08 |
| Log $P_{o/w}$ (SILICOS-IT) | 2.72 | -0.98 |
| Consensus Log $P_{o/w}$ | 4.35 | -0.22 |
| | Water Solubility | |
| Log $S$ (SILICOS-IT) | -8.73 | -2.94 |
| class | Poorly soluble | Soluble |
| Solubility | 1.62e-06 mg/ml ; 1.85e-09 mol/l | 5.07e-01 mg/ml ; 1.14e-03 mol/l |
| | Pharmacokinetics | |
| Gatrointestinal absorption | Low | Low |
| BBB permanent | No | No |
| P-gp substrate | Yes | Yes |
| CYP1A2 inhibitor | No | No |
| CP2C19 inhibitor | No | No |
| Log $K_p$ (skin permeation) | -7.14 cm/s | -8.63 cm/s |
| | Drug Likeness | |
| Lipinski Rule | No; 2 violations: MW>500, NorO>10 | Yes; 1 violation: NHorOH>5 |
| Ghose Filter | No; 4 violations: MW>480, WLOGP>5.6, MR>130, #atoms>70 | No; 1 violation: WLOGP<-0.4 |
| Veber (GSK) Rule | No; 1 violation: TPSA>140 | No; 1 violation: TPSA>140 |
| Egan (phatmacial) Filter | No; 1 violation: TPSA>131.6 | No; 1 violation: TPSA>131.6 |
| Muegge (Bayer) Filter | No; 4 violations: MW>600, XLOGP3>5, TPSA>150, H-acc>10 | No; 2 violations: TPSA>150, H-don>5 |
| Bioavailability (Abbott) Score | 0.17 | 0.11 |
| | Medicinal Chemistry | |
| PAINS (Pan Assey Interference Structures) | 0 alert | 0 alert |
| Brenk | 1 alert: isolated_alkene | 1 alert: michael_acceptor_4 |
| Leadlikeness | No; 3 violations: MW>350, Rotors>7, XLOGP3>3.5 | No; 1 violation: MW>350 |
| Synthetic accessibility | 10.00 | 5.25 |

For the first experienced inhibitor ivermectin is docked with 3CL$^{pro}$, 6LU7 protein. Based on molecular docking ivermectin: 6LU7 complex revealed 9 different poses. For finding out the best pose for the ligand and receptor complex formation, molecular docking simulation follows certain rules. The pose with highest negative values of binding energy, a greater number of hydrogen bonds and lowest value of dreiding energy and dipole moment considered as the best one. For ivermectin: 6LU7 complex, we have observed pose 3 is the better interacted position for ligand: protein complex with the binding affinity of -6.9 kcal/mol. We have also computed the dreiding energy of different poses, in order to confirm the most excellent docked site. The dreiding energy (6,298.99) becomes minimum for the best docked 3 pose (Table 2).

To confirm the better interaction between ivermectin and 6LU7, we have calculated the inhibition constant ($k_i$). It normally indicates how potent drugs inhibitors are towards protein. The inhibition constant can be calculated using the following equation:

$$k_i = e^{\frac{\Delta G}{RT}} \quad \ldots\ldots\ldots\ldots\ldots\ldots\ldots\ldots\ldots\ldots(5)$$

where G is binding affinity, R is universal constant and T is the room temperature (298 K).

For the best docked 3 pose of ivermectin: 6LU7 complex, the obtained value of ki as 8.7 X 10$^{-6}$M which proves the strong attraction of ivermectin towards protein 6LU7 (Table 2). The strong interaction for best docked pose (3) was further confirmed by the number of weak non-bonded hydrogen bonded interactions and hydrophobic interactions present between protein: ligand complex structure. "Hydrogen bonding and hydrophobic interactions" always stabilize the ligands at the target protein site [62]. We have observed the presence of intermolecular hydrogen bonds and hydrophobic interaction between protein and ligand. For best poses of ivermectin: protein complex, the donor–acceptor surface and different possible interactions in 3D and 2D view are shown in Figure 3 a.

Same molecular docking approach has been done for doxycycline ligand with protein 6LU7. In terms of their different parameters (binding affinity value, dreiding energy, dipole moment, inhibition constants, number of hydrogen bonds, hydrophobic bonds etc.), we have identified the best possible ligand: protein docked pose position (Table 2). For doxycycline: protein complex, pose 7 is the better interacted position with the binding affinity of -6.4 kcal/mol, dreiding energy; 6,063.5, dipole moment; 6.104 Debye, inhibition constant; 2.0 X 10$^{-5}$ M and 7 number of hydrogen bonds (Table 2). Best pose of the donor–acceptor surface with their possible hydrogen bonding and hydrophobic interactions 3D and

2D view are shown in Figure 3b. Our result shows that out of two possible ligand drug structures, ivermectin represents the best potentiality to inhibit with the SARS 3CL$^{pro}$ (6LU7) by its best docking affinity compared to the doxycycline. Good binding mode of interactions of ivermectin: 6LU7 complex also verified by its less binding energy, minimum inhibition constant value as compared to doxycycline. Both the drug molecules showed good stability as a complex with the targeted protein. These drug molecules also satisfy the required drug likeness properties according to Ro5, Veber etc. rules, polar surface areas and logP values.

**Table 2.** Interaction factor for Ivermectin and Doxycycline with receptor protein (6LU7).

| Protein | Binding affinity (kcal/mol) | Hydrogen bonded interaction (donor: acceptor, distance in A) [Type of bond] | Dipole moment (ligand) Debye | Dreiding energy (protein+ligand) | Inhibition constant (M) |
|---|---|---|---|---|---|
| **docking (Ivermectin)** | | | | | |
| **6LU7** | -6.9 | (A:THR25:HG1-:UNL1:O, 2.87667) [Conventional Hydrogen Bond]<br>(A:THR26:HN-:UNL1:O, 2.03169) [Conventional Hydrogen Bond]<br>(A:ASN142:HD22-:UNL1:O,2.79324)[Conventional Hydrogen Bond]<br>(:UNL1:H-A:THR26:O, 2.45352) [Conventional Hydrogen Bond]<br>(:UNL1:H-A:THR26:O, 2.13735) [Conventional Hydrogen Bond]<br>(A:THR25:CA-UNL1:O, 3.40238) [Carbon atom Hydrogen Bond]<br>(A:PRO168:CA-UNL1:O, 3.78628) [Carbon atom Hydrogen Bond] | 5.830 | 6,298.99 | 8.7 X 10$^{-6}$ |
| **docking (Doxycycline)** | | | | | |
| **6LU7** | -6.4 | (A:ASN142:HD22-:UNL1:O, 3.03586)[Conventional Hydrogen Bond]<br>(:UNL1:H-A:GLU166:O, 2.0088) [Conventional Hydrogen Bond]<br>(:UNL1:H-A:GLN189:OE1, 2.12427) [Conventional Hydrogen Bond]<br>(:UNL1:H-A:ASN142:OD1, 2.23539) [Conventional Hydrogen Bond]<br>(:UNL1:H-:UNL1:O, 2.78645) [Conventional Hydrogen Bond]<br>(:UNL1:H-:UNL1:O, 2.48006) [Conventional Hydrogen Bond]<br>(:UNL1:C-A:HIS163:NE2, 3.59178) [Carbon atom Hydrogen Bond] | 6.104 | 6,063.5 | 2.0 X 10$^{-5}$ |
| **Sequential docking (Ivermectin+Doxycycline)** | | | | | |
| **6LU7** | -7.4 | (A:GLY143:HN-:UNL1:O, 2.122) [Conventional Hydrogen Bond]<br>(A:GLY143:HN-:UNL1:O, 2.86002) [Conventional Hydrogen Bond]<br>(A:SER144:HN-:UNL1:O2.32648) [Conventional Hydrogen Bond]<br>(A:SER144:HG-:UNL1:O2.1576) [Conventional Hydrogen Bond]<br>(A:CYS145:HN-:UNL1:O,2.57732) [Conventional Hydrogen Bond]<br>(A:GLU166:HN-:UNL1:O, 2.23187) [Conventional Hydrogen Bond]<br>(UNL1:H- A:LEU141:O , 2.4969) [Conventional Hydrogen Bond]<br>(:UNL1:H-:UNL1:O, 1.95212) [Conventional Hydrogen Bond]<br>(:UNL1:C-A:ASN142:OD1, 3.4013) [Conventional Hydrogen Bond]<br>(:UNL1:C-A:HIS41:NE2, 3.42481) [Conventional Hydrogen Bond]<br>(:UNL1:C-A:GLN189:OE1, 3.5951) [Carbon atom Hydrogen Bond] | 2.237 | 6,408.28 | 3.7 X 10$^{-6}$ |

**Figure 3.** Donor: acceptor surface and possible types of interactions in best pose structures obtained from molecular docking for **a)** ivermectin: 6LU7 **b)** doxycycline: 6LU7 complex.

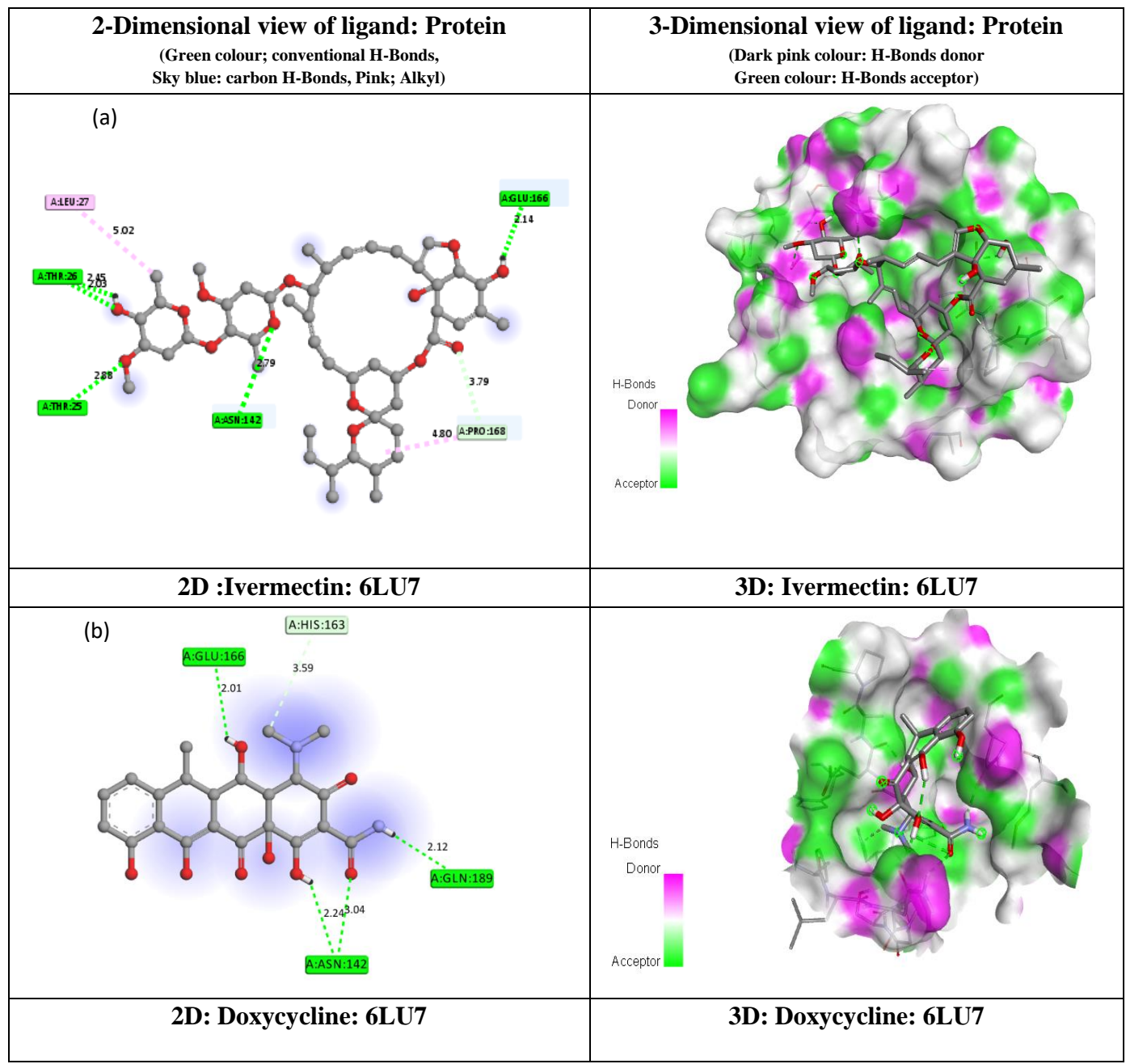

## 3.2. Sequential docking of two drugs against SARS-CoV-2 protease

Individually, ivermectin and doxycycline drugs showed a good binding energy of -6.9 kcal/mol and -6.4 kcal/mol, respectively. The docked ligand molecules with the protease 3CL$^{pro}$ (6LU7) are shown in Figure 4a,b. The possible hydrophobic interactions and hydrogen bond between 6LU7of the two considered drugs obtained with individual docking are presented in Table 1. We have performed sequential docking for checking the interaction of combinational drugs (two or more than two drugs mixed to form a single drug) and the target protein. In the usual docking procedure we docked a single ligand with the receptor protein. However, with the help of sequential docking it is possible to dock more than two ligands simultaneously. This is helpful for detecting allosteric (place on protein where ligand that is not a substrate may bind) binding site. In the present work we have also checked the interaction of a combination of drugs (ivermectin+doxycycline) with the 6LU7 protein. Sequential docking of two drugs simultaneously with the 6LU7 protein showed a significant enhancement in the binding energy to -7.4 kcal/mol (Figure 4 c). In Figure 4, the red circle indicates the binding drug site with their binding energies respectively. The two most suitable nearest poses which validate the best pose 1 structure for ivermectin+doxycycline: 6LU7 complex is shown in supporting document 1 (SD1). Since sequential docking of ivermectin and doxycycline drugs with 6LU7 shows the better possibility of inhibition we have further studied the applicability of combination of these drugs as a potential drug by using MD simulation approach.

**Figure 4.** Binding energies of **(A)** ivermectin: 6LU7 **(B)** doxycycline: 6LU7 **(C)** ivermectin+doxycycline: 6LU7 complex. The drug binding site is indicated by a red circle with their respective binding energies.

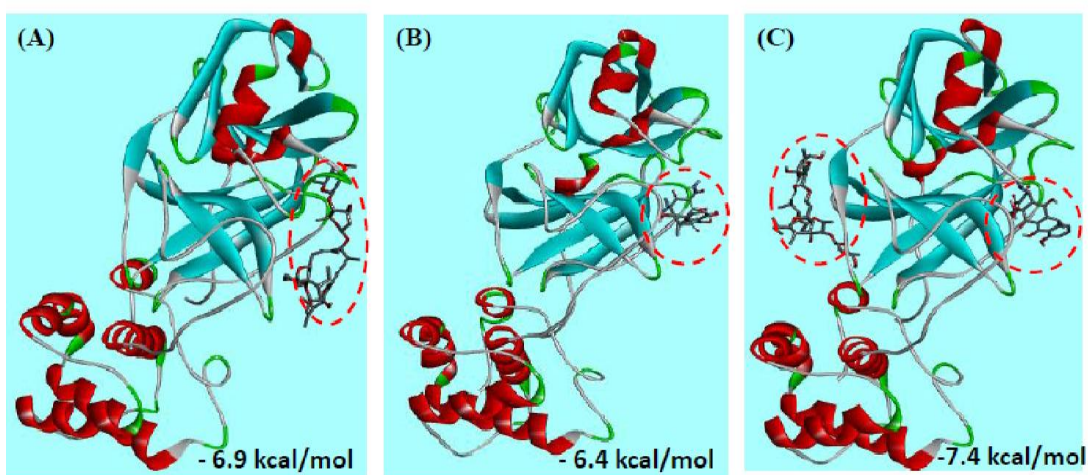

The stability of the particular complex is directly proportional to the number of nonbonded interactions. Larger the number of nonbonded interactions the more possibility of formation of complex structure. Maximum number of conventional hydrogen bonds and carbon hydrogen bonds (weak interactions) were observed for pose 1 of docked structure between ivermectin+doxycycline: 6LU7 complex (Figure 3).

## *3.3. MD simulation analysis*

MD simulation can simulate in picoseconds/nanoseconds or further finer temporal stead-fastness [63]. The MD simulation force field plays an important role for estimating the forces within the molecule (intramolecular force) and between two molecules (intermolecular force). These intermolecular and intramolecular forces used to calculate the potential energy of the molecules. The total energy of the system is given as the sum of bonded and non-bonded energy and given as below:

$$E_{total} = E_{bonded} + E_{non\ bonded} \ldots\ldots\ldots\ldots\ldots\ldots\ldots\ldots\ldots (6)$$

$$E_{bonded} = E_{bond} + E_{angle} + E_{dihedral} \ldots\ldots\ldots\ldots\ldots\ldots\ldots\ldots (7)$$

$$E_{non\ bonded} = E_{hydrogen\ bond} + E_{electrostatic} + E_{vander\ waals} \ldots\ldots\ldots (8)$$

$$E_{electrostatic} = E_{coulombic} + E_{lenard\ Jones} \ldots\ldots\ldots\ldots\ldots\ldots\ldots (9)$$

These equations show that the bonded energy is the combination of bond, angle and dihedral energies while nonbonded energy is the combination of hydrogen bond, electrostatic and van der waals energies (eq. 7, 8).

To analyze the stability of the studied structure, MD simulation of the complexes (ivermectin: 6LU7, doxycycline: 6LU7, ivermectin+doxycycline: 6LU7) have been studied for the period of 10000 ps to 100 ps. For MD simulation, first we have to make all the structure energetically optimized (the potential energy should be minimum and negative with a maximum force value). Figure 5 represents energetically minimized protein and complex systems. We have obtained steady convergence of potential energy for all the cases. The comparison of the potential energy (Epot) of the stable structure of bare 6LU7 protein and in drugs: 6LU7 complex has been done carefully. In the bare state 6LU7 has Epot of $-1.27 \times 10^6 \pm 56.7$ kJ/mol, while the complex ivermectin: 6LU7, doxycycline: 6LU7 and ivermectin+doxycycline: 6LU7 has

an average Epot of –2.55661X10$^5$±11kJ/mol, –2.54672X10$^5$±14 kJ/mol, –2.48721X10$^5$ ±53 kJ/mol respectively (Table 3). Now all the structures having their lowest Epot values are ready for MD simulation.

**Figure 5:** Potential energy surface for optimized geometry of bare 6LU7, ivermectin: 6LU7 complex, doxycycline: 6LU7 complex and ivermectin+doxycycline: 6LU7 complex.

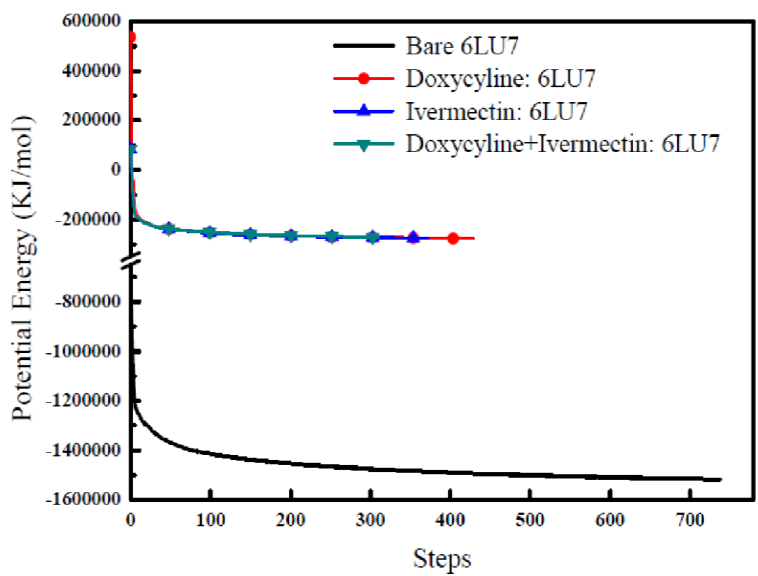

**Figure 6:** Temperature progression data for bare 6LU7, ivermectin: 6LU7, doxycycline: 6LU7 and ivermectin+doxycycline: 6LU7 complex in water environment in GROMOS and Charm 36 force fields.

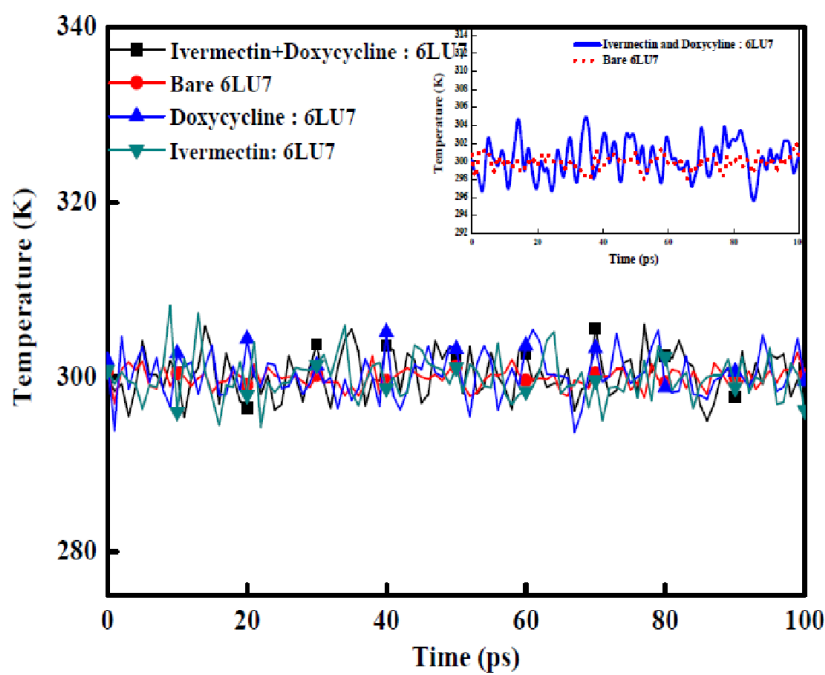

To stabilized different parameters (temperature (T), pressure (P), density (D), volume (V) etc.) within a time scale of 100 ps to 10000 ps, we have further check the optimized drugs: 6LU7 protein structures equilibrated by NVT and NPT ensembles. It is observed that over the period of 100 ps time trajectory the temperature of the complex rapidly reached the stable at 300 K (room temperature) value (Figure 6). This temperature stability is maintained throughout the process. The temperature, pressure and density values of the system were also observed to be very stable over the period of time trajectory 100 ps (SD2, SD3). This concludes that the system is well equilibrated and prepared for MD simulation.

The compactness of the system with respect to time of a bare protein and protein: ligand complex can be measured with the help of radius of gyration ($R_g$) [64]. Normally for the stably folded protein structures the values of $R_g$ keeps a relatively steady for full time scale [65]. Whereas the $R_g$ values for the unfolded protein keeps changing for full time scale. Less compactness in the structures and high compactness with more stability exhibit a low and high $R_g$ value respectively. In the present paper we have observed the bare protein (6LU7) has $R_g$ in between 2.25 nm – 2.26 nm with an average value of 2.225 nm (SD 4, Table 3). Almost similar variation is observed with the proposed dugs: 6LU7 complex (SD 4). This shows high compactness with more stability in the protein and drugs complexes (SD 4).

Further to validate the applicability of ivermectin, doxycycline and ivermectin+doxycycline ligands as proposed drug for COVID-19, we have simulated the SASA. SASA measures the area of exposure of the receptor to the solvents. The higher value of SASA indicates that the drug is more inserted into the water whereas, lower value represents that more drug is covered by the protein, which represents better complexation. In the present work, we have obtained the SASA value in the range of 19–26 $nm^2$ for bare protein with the mean value 22 $nm^2$ (SD 5, Table 3). Similarly, for all the proposed drug and protein 6LU7 complex the mean value of SASA is 9 $nm^2$. The low computed values of SASA observed for all drugs: protein complex shows that drug binding with the receptor protein increases the exposure of complexes to the protein (SD 5). Which validates the best complexation possibility.

**Figure 7:** Hydrogen bond number for optimized geometry of **a)** ivermectin: 6LU7 complex, **b)** doxycycline: 6LU7 complex, and **c)** ivermectin+doxycycline: 6LU7 complex.

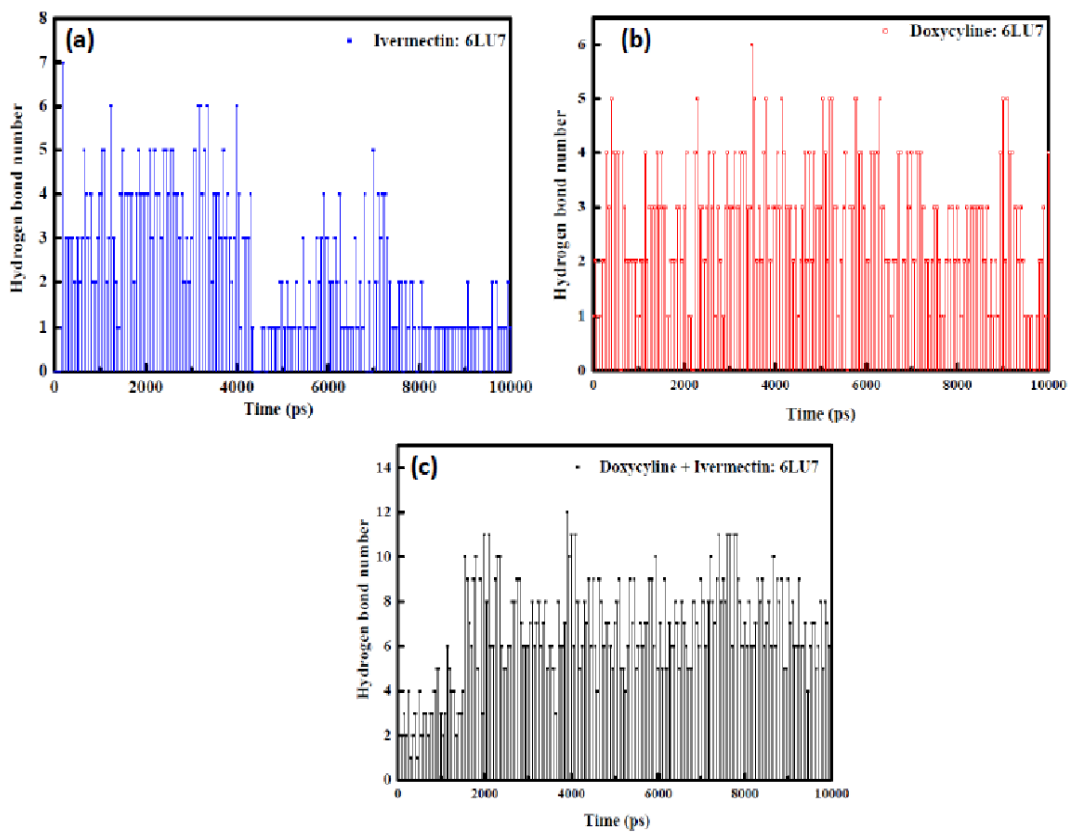

Intermolecular hydrogen bonding plays a significant role to get an idea about the binding strength between protein and drug. Ivermectin has a stable range of intermolecular hydrogen bonding with protein between 0 to 7 with an average value 3.5 in throughout the whole simulation process (Figure 7, Table 3). Doxycycline has a range of intermolecular hydrogen bonding with protein between 0 to 6 in throughout the whole simulation process with an average value of 3. However, the combination of both the drugs (ivermectin+doxycycline) has the highest stable range of intermolecular hydrogen bonding with protein between 0 to 12 with an average value 7 (Table 3). The intermolecular hydrogen bond number computed through MD simulation also perfectly matches with the docking results. This result clearly indicates that there is no conformational change around the probe drug systems in the binding site throughout the simulation process (Figure 7). The appearance of larger intermolecular hydrogen bonding in combination phase of ivermectin+doxycycline with the target protein 6LU7 validates best binding phase compared to single phase binding with receptor protein.

**Figure 8.** Root mean square deviation (RMSD) of receptor protein 6LU7 in its bare state, ivermectin: 6LU7 complex, doxycycline: 6LU7 and ivermectin+doxycycline: 6LU7 complex **a)** 3D view and **b)** 2D view up to 10ns.

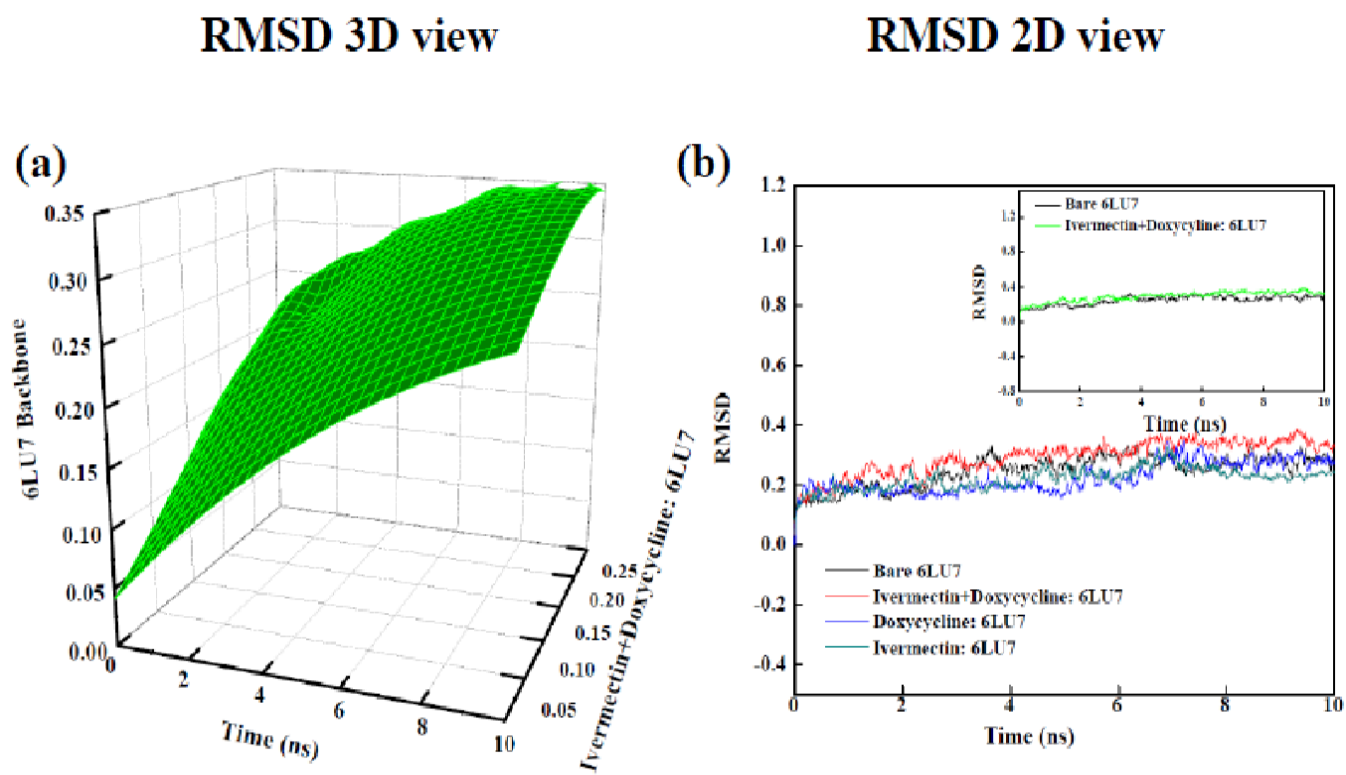

RMSD corresponds to any change in the conformational stability of the protein: drug complex and in the protein dynamics. RMSD of the free protein and protein: ligand complex have been simulated to 100 ns by using MD simulations. RMSD and RMSF have been measured by using the GROMACS module at an interval of 1 ns. RMSD variation of bare 6LU7 protein lies in the range from 0.08 to 0.16Å. Ivermectin: 6LU7, doxycycline: 6LU7, ivermectin+doxycycline: 6LU7 complex, also ranges RMSD values from 0.08 to 0.16 Å (Table 3). The RMSD value for complexes exactly matches with the bare protein. This provided a suitable basis for our study by the better stability with the probe drugs. Figure 8 represents the 2D and 3D view of RMSD values of Cα atoms of the bare protein and protein: ligand complex individually at various nanoseconds. The RMSD graph of all three ligands showed stability during the simulations (Figure

8). We have observed all the complexes are stable and no deviations of RMSD values were found throughout the simulations.

**Figure 9.** Graph of root mean square fluctuations (RMSF) of 6LU7 in its bare state and in ivermectin+doxycycline: 6LU7 complex.

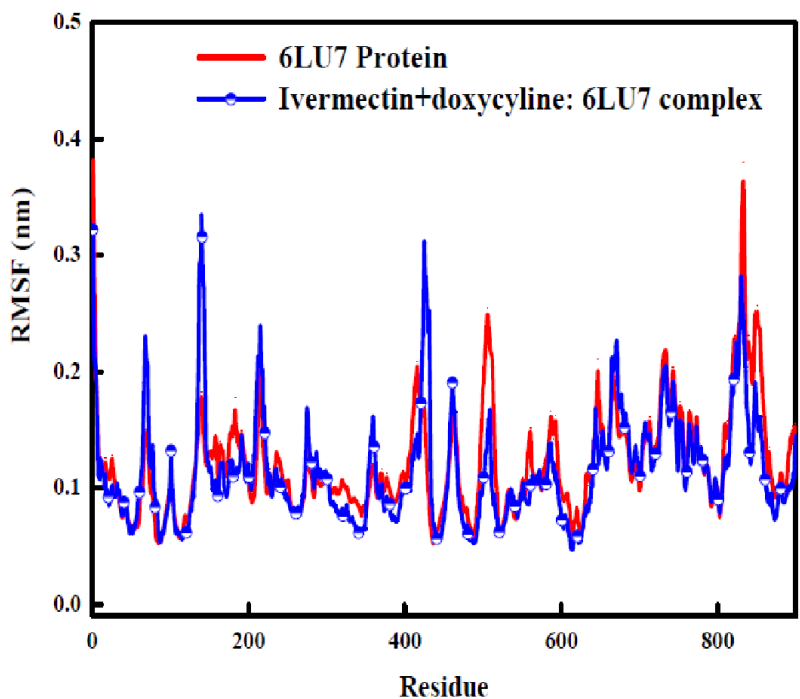

For all amino acid residues with respect to Cα atom RMSF have been simulated. RMSF plot for 6LU7 in its bare state and ivermectin+doxycycline: 6LU7 complex have been shown in Figure 9, which depicts the fluctuations at the residue level. Residue fluctuation profile for both the cases shows a similar trend having an average RMSF value of 0.15 Å, which indicates that binding of both the drugs to the 6LU7 protein had no key effect on the flexibility of the protein and was quite stable.

**Table 3.** MD simulation output parameters of 6LU7 in its bare state without any ligand and in the ivermectin+doxycycline: 6LU7 complex.

| Serial No. | Parameter | Bare 6LU7 | | Ivermectin: 6LU7 complex | | Doxycycline: 6LU7 complex | | Ivermectin+Doxycycline: 6LU7 complex | |
|---|---|---|---|---|---|---|---|---|---|
| | | Mean | Range | Mean | Range | Mean | Range | Mean | Range |
| 1. | SR Coulombic Interaction Energy (kJ/mol) | NA | NA | -70.1484 ±14 | -136--21 | -89.5844 ±2.9 | -74--106 | -84.9295 ±13 | -90--40 |
| 2. | SR Lennard-Jones Interaction Energy (kJ/mol) | NA | NA | -83.803 ±17 | -135--49 | -95.8948 ±4.5 | -87-113 | -125.189±3.1 | -135--111 |
| 3. | RMSD (nm) | 0.12 | 0.08–0.16 | 0.12 | 0.08–0.16 | 0.12 | 0.08–0.16 | 0.12 | 0.08–0.16 |
| 4. | Inter H-Bonds | NA | NA | 3.5 | 0-7 | 3 | 0-6 | 7 | 0-12 |
| 5. | Radius of gyration | 2.25 ± 0.01 | 2.25–2.26 | 2.91 | 2.91-2.93 | 2.25 | 2.25–2.26 | 2.91 | 2.91-2.93 |
| 6. | SASA (nm$^2$) | 22 | 19–26 | 9 | 4-14 | 9 | 4-14 | 9 | 4-14 |
| 7. | Potential Energy (kJ/mol) | -1.27X10$^6$ ± 56.7 | -7.3 X10$^5$ - -1.3 X10$^6$ | -2.55661 X 10$^5$ ±11 | -5.3 X10$^4$ - -2.77 X10$^5$ | -2.54672 X 10$^5$ ±14 | -5.3 X10$^4$ - -2.77 X10$^5$ | -2.48721 X 10$^5$ ±53 | -5.3 X10$^4$ - -2.77 X10$^5$ |
| 8. | Binding energy($\Delta$G)(KJ/mol) | NA | NA | -8.718 ± 25.676 | NA | -6.677 ± 41.724 | NA | -10.603 ± 41.086 | NA |
| 9. | Van der Waal Energy($\Delta E_{vdw}$) (KJ/mol) | NA | NA | -0.065 ±0.088 | NA | -23.492 ± 49.768 | NA | -0.542 ± 7.940 | NA |
| 10. | Electrostatic Energy($\Delta E_{elec}$)(KJ/mol) | NA | NA | -0.324 ±0.704 | NA | -8.506 ±19.752 | NA | -0.483 ± 3.815 | NA |
| 11. | Polar Solvation Energy(($\Delta E_{polar}$)(KJ/mol) | NA | NA | -8.527 ±25.583 | NA | 27.758 ± 43.697 | NA | -9.737 ±41.527 | NA |
| 12. | SASA Energy ($\Delta E_{apolar}$)(KJ/mol) | NA | NA | 0.197 ±2.426 | NA | -2.437 ± 5.567 | NA | 0.159 ± 3.100 | NA |

The short-range nonbonded interaction energy (Coulombic short range protein: ligand interaction energy terms and Lennard Jones short range protein: ligand interaction energy terms) quantify the strength of the interaction between probe drugs and protein. Addition of Coulombic interaction energy and Lennard Jones interaction energy provides the total interaction energy. Figure 10 a,b shows the contour map and 3D graph of obtained total interaction energy for the ivermectin+doxycycline: 6LU7 complex. The average Coulombic interaction energy for ivermectin+doxycycline: 6LU7 complex comes out –84.9295±13 kJ/mol whereas the average Lennard-Jones interaction energy is –125.189±3.1 kJ/mol (Table 3). Table 3 represents all the Coulombic interaction energy and Lennard Jones interaction energy for individual drugs: protein complex and combination of drugs: protein complex. The comparison suggests that for all the complex formation, short-range Lennard-Jones has shown stronger effect on binding affinity than the short range Coulombic interaction energy.

**Figure 10:** For ivermectin+doxycycline: 6LU7 complex **a)** contour plot of coulombic interaction energy and Lenard Jones interaction energy **b)** 3D representation of coulombic interaction energy and Lenard Jones interaction energy with respect to the time trajectory (0 to 10000 ps).

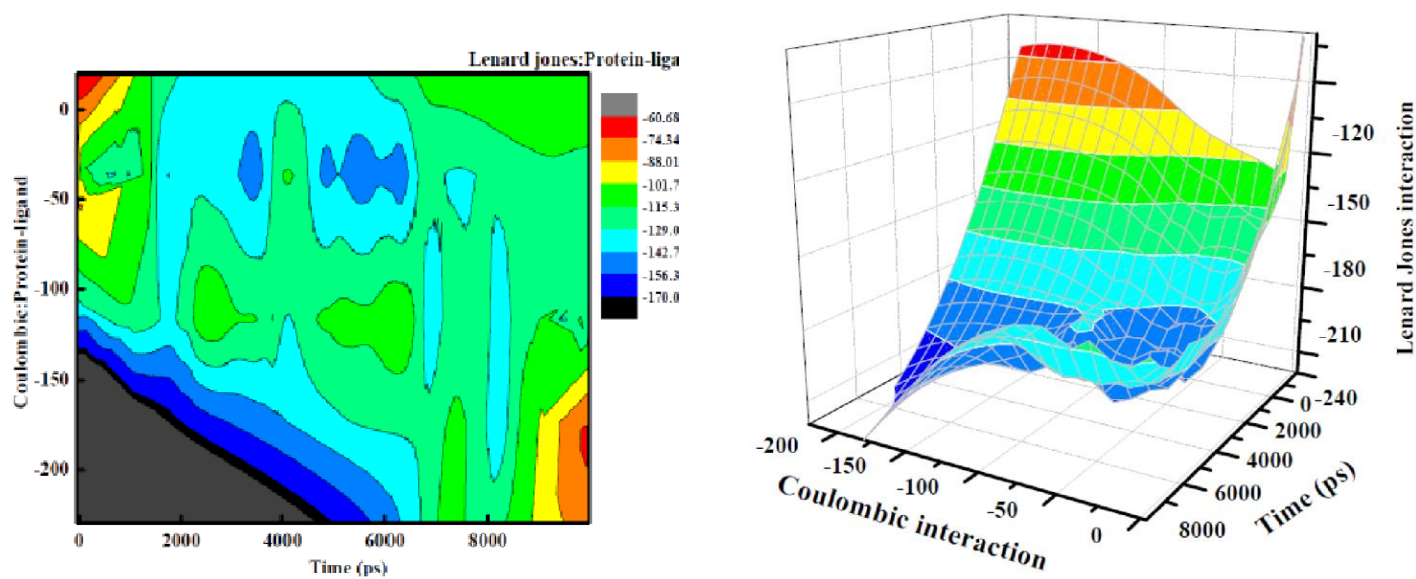

For the complex formation ΔG indicates the non-bonded interaction energies which is the sum of comprehensive energies of individual components while the binding energy through molecular docking provides only binding energy of the complex formation. Based on different quantum simulation techniques, there are a number of research works going on to check the stability of varieties of complex configurations based on interaction energies [66,67] . Figure 11 represents the ΔG values for ivermectin: 6LU7, doxycycline: 6LU7 and ivermectin+doxycycline: 6LU7 complex with respect to the time trajectory 0 ps to 10000 ps. The observed ΔG values for ivermectin+doxycycline: 6LU7 complex is the lowest (– 10.603 ± 41.086 kJ/mol) in comparison of other complexes (ΔG for ivermectin –8.718±25.676, ΔG for doxycycline –8.718±25.676)(Table 3). This clearly indicates that ivermectin and doxycycline makes better complexation with the SARS-CoV-2 protein but the combination of these two drugs can make impressively best stable complex formation with receptor protein 6LU7.

**Figure 11:** Total binding energy with respect to the time trajectory (0 to 10000 ps) for ivermectin: 6LU7 complex, doxycycline: 6LU7 complex and ivermectin+doxycycline: 6LU7 complex.

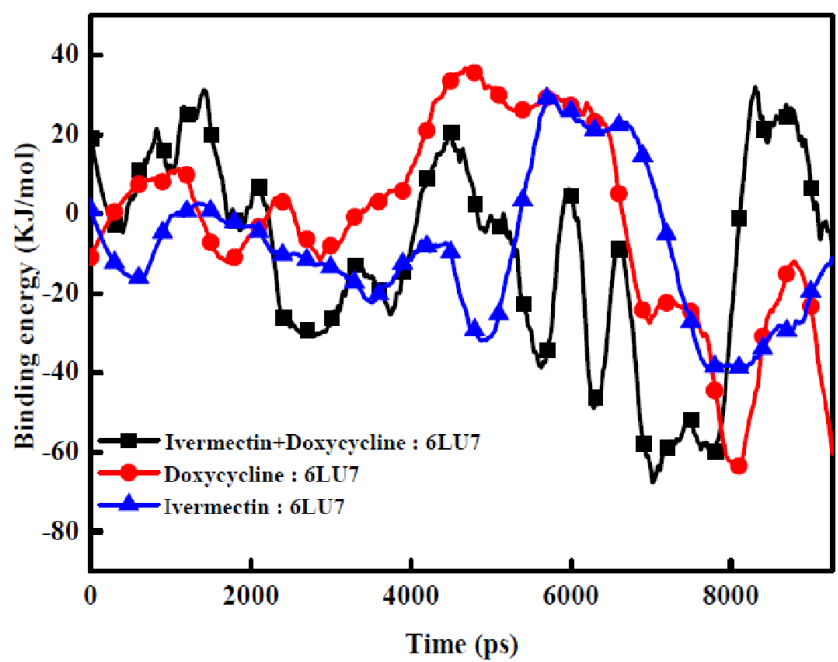

## 4. Conclusions

In conclusion, two drugs (ivermectin and doxycycline) were tested as potential inhibitors for COVID-19 main protease 6LU7 via molecular docking. A strong inhibitory possibility of proposed drugs for SARS-CoV-2 protease 3CL$^{pro}$ was verified by physiochemical, pharmacokinetics, drug likeness, and medicinal chemistry properties by using ADME analysis. From docked compounds, we have proposed that ivermectin and doxycycline demonstrated high binding affinity to the 6LU7 and their combined docking increases the binding affinity on COVID-19 main protease. Strong binding affinity, lowest inhibition constant and existence of hydrogen bonded interaction established the better stability of ivermectin+doxycycline: 6LU7 complex structure. Further studies also conducted on these compounds using MD simulations in order to get more reliable data. Many thermodynamic parameters ($E_{pot}$, T, V, D, $R_g$, SASA energy) obtained by MD simulation also validated the complexation between ivermectin+doxycycline and 6LU7. The backbone of the complex and free 6LU7 protein illustrate similar RMSD and RMSF, which demonstrate the stability of the binding of drugs and protein. MD analyses have also confirmed the complexation between proposed drug and 6LU7 protein by the lower values of binding energy. All simulated results establish that combination of drugs is a stronger candidate as a potential inhibitor for SARS-CoV-2 than considering each drug separately. Our present in silico study would provide a new approach to the researchers working in the field of new drug finding against SARS-CoV-2. However, a proper in-vivo and in-vitro rigorous research works are to be performed for the validation of our simulation work so that our recommended combination drug may be considered as a promising candidate for the drug design against COVID-19.